\documentclass{ws-procs975x65}
\usepackage{epsfig,amsmath,amssymb,psfrag}
\newcommand{\Cl}{{C_\ell}}
\newcommand{\Chat}{{\hat{C}_\ell}}
\newcommand{\fsky}{f_{\rm{sky}}}

\newcommand{\alt}{\sim<}
\newcommand{\agt}{\sim>}
\newcommand{\muK}{{\mu{\rm K}}}
\newcommand{\chieff}{\chi^2_{\rm{eff}}}

\begin{document}

\title{Cosmological parameters and the WMAP data}
\author{Antony Lewis}

\address{Astronomy Dept., Harvard University, 60 Garden St, Cambridge, MA 02138, USA\\
CITA, University of Toronto, 60 St. George St, Toronto, ON M5S 3H8, Canada\\E-mail: Antony@AntonyLewis.com}

\maketitle

\abstracts{
I discuss whether the standard cosmological
  models fit the WMAP data well enough to justify parameter estimation with
  standard assumptions.
The observed quadrupole is low (but has significant
  foreground uncertainty) and drives weak evidence for theoretical 
models predicting low values, such as models with a running spectral
  index. Other more seriously
  outlying points of the WMAP power spectrum appear not to fit the
  expectations of simple Gaussian models very well. The effective temperature
  chi-squared is however acceptable on large scales. There also appears to be
  evidence for an anisotropic distribution of power, which 
  taken together with the other points may indicate that either there
  is a problem with the WMAP data or that standard cosmological models are incorrect.
 These issues should be clarified before
  cosmological parameter extraction for the usual standard models can
  be trusted, and hint that maybe the CMB is more interesting than we imagined.
  I also discuss various systematic and analysis issues, and comment
  on various oddities
in the  publicly available first year WMAP data and code.
}

\section{Introduction}

The recent results of the Wilkinson Microwave Anisotropy Probe
(WMAP)\cite{Hinshaw:2003ex} 
provide full-sky maps of the CMB anisotropy together
with various foregrounds and noise. Taken at face value the anisotropy power
spectrum can be used to tightly constrain various combinations of
cosmological parameters. Partial parameter degeneracies can be broken by
including additional data from other sources, giving good constraints on many
parameters individually\cite{Spergel:2003cb,Contaldi:2003hi}. However 
data from Lyman-$\alpha$ forest is currently plagued by
systematic issues (for example see Ref.~\refcite{Seljak:2003jg}), so
joint constraints including only statistical errors should not be
taken too seriously. Even cosmic shear results, 
 which in principle are a clean
probe of the total matter distribution
and should give robust joint parameter
constraints\cite{Contaldi:2003hi}, are currently subject to various
observational systematics and uncertainties than can be hard to
quantify. Constraints on the matter power spectrum from
2dF\cite{Percival:2001hw} are sensitive to how the bias is modelled,
with conservative results that are marginalized over the bias\cite{Bridle:2003sa} giving
significantly higher results for the matter density than when more
complicated modelling is applied\cite{Verde:2003ey}. Having said this, 
there is a remarkable agreement between the cosmological
parameters inferred from many of these data sets and different
CMB power spectrum measurements, indicating that the
basic flat $\Lambda$CDM model with approximately power-law isotropic adiabatic
Gaussian primordial fluctuations may be on the right track.

In principle the CMB anisotropy measurement on large scales should be
the cleanest and most direct probe of primordial physics. Usually the
sky maps are compressed into power spectra for the purpose of
likelihood evaluation, which should be nearly optimal if the
fluctuations are indeed isotropic and Gaussian, and the power spectrum
and likelihoods are evaluated consistently. Given a function for
computing the likelihood from a theoretical power spectrum, computing
the posterior parameter constraints is straightforward using the
publicly available CosmoMC code\cite{Lewis:2002ah}\footnote{
{\sf http://cosmologist.info/cosmomc/}}. However as discussed
by Ref.~\refcite{Spergel:2003cb} and other authors the WMAP power spectrum has some
unexpected features, and seems to have a rather high $\chieff$ fit to
standard models. I discuss these features in more detail below, where
bulleted points are of a technical or trivial nature and may be skipped.

\section{Temperature power spectrum}

\begin{figure}[ht]
\psfrag{l}[][][0.7]{$\ell$}
\psfrag{Cl}[][][0.7]{$\ell(\ell+1) C_\ell/(2\pi \mu \text{K}^2)$}
\psfig{figure=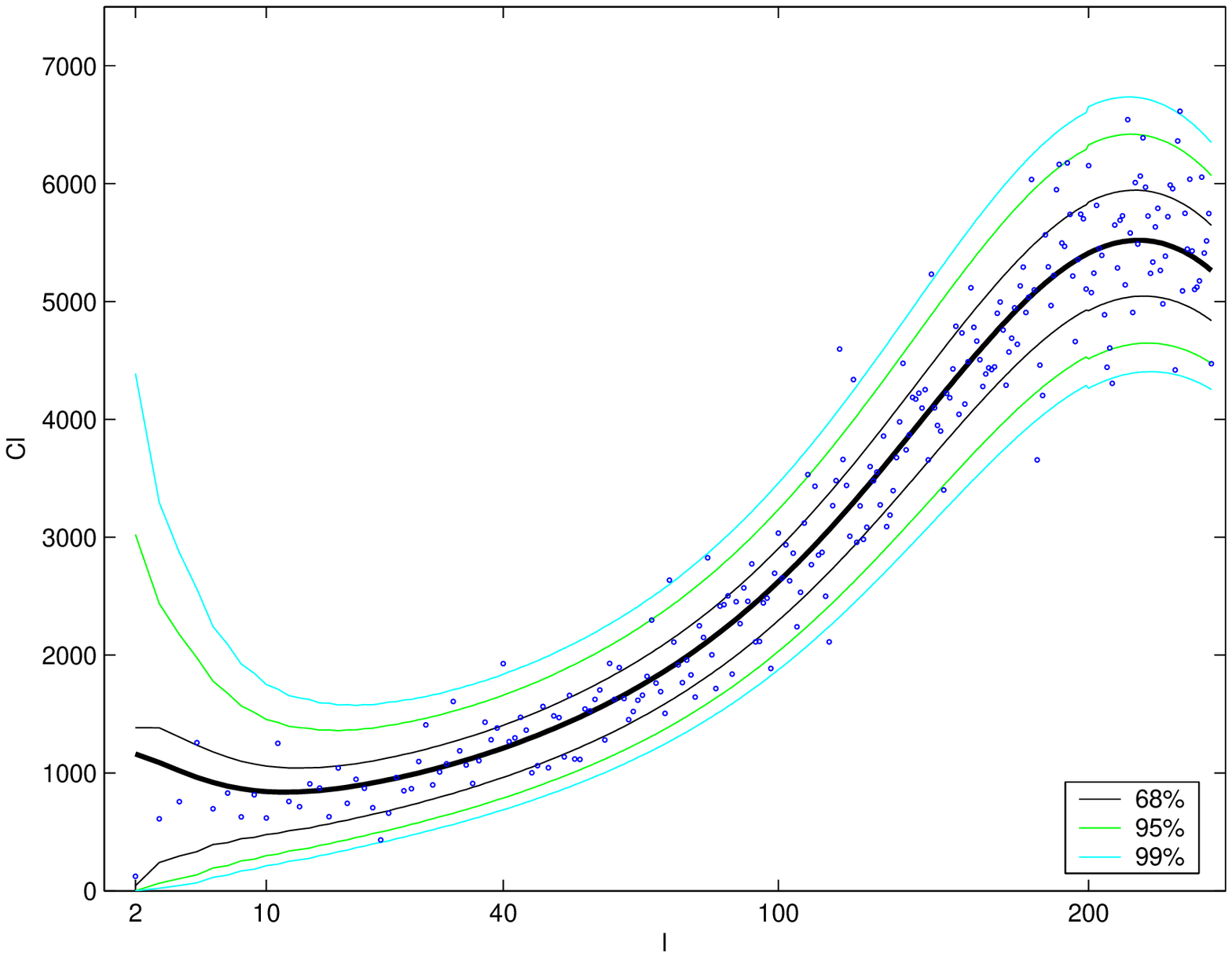, height=2in}
\psfig{figure=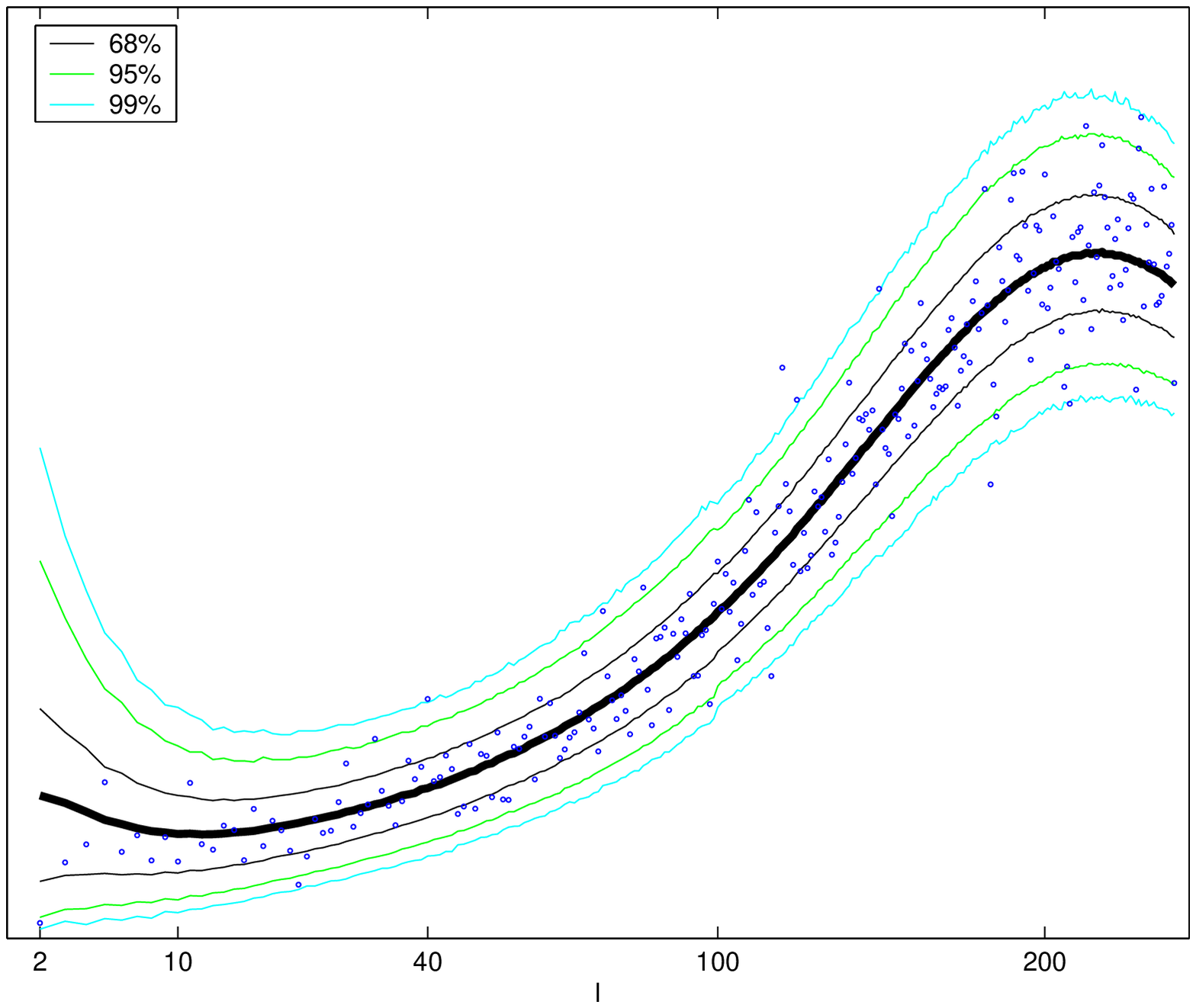, height = 2in}
\caption{The WMAP power spectrum at low $\ell$ (where it is signal
  dominated). The thick line shows a theoretical spectrum (the mean
  of each value over many realizations). The points are the WMAP
  $\Chat$ estimates. The other lines on the left plot are
  isoprobability contours enclosing $68\%$, $95\%$ and
  $99\%$ for the appropriate $\chi^2$ distribution (no noise), lines on
  the right plot are from pseudo-$\Cl$ simulations with noise, where
  $32\%$, $5\%$ and $1\%$  of $\Chat$ realizations are equi-distributed below and above the
  different lines. 
  This assumes the plotted theoretical model is correct.
\label{Cl}}
\end{figure}

In Fig.~\ref{Cl} I have plotted the WMAP temperature $\Chat$ estimates at
$\ell<250$ where the uncertainty from the noise is not too large. On
the full sky and with no noise the maximum likelihood estimators are given by $\Chat = \sum_m
|a_{lm}|^2/(2l+1)$ where the $a_{lm}$ are the coefficients of the
spherical harmonic expansion of the CMB temperature on the sky. With cuts
around foregrounds like the galaxy and in the presence of noise things
are not quite so simple. The values plotted are pseudo-$\Cl$
estimators, which are suboptimal but simple to compute as described in
Refs.~\refcite{Wandelt00,Hinshaw:2003ex}. For a given theoretical model
the probability distribution for the $\Chat$ estimators in any
particular realization of the sky is something like a
$\chi^2$ with $(2l+1)\fsky^2$ degrees of freedom\cite{Verde:2003ey} (where $\fsky$ is the
fraction of sky observed). The most likely values of the
estimators are therefore actually \emph{below} the mean value (which is the
variance of the $a_{lm}$s predicted by the theoretical model). 

The limits in Fig.~\ref{Cl} are approximate and depend on the
theoretical model. However
there still appear to be a number number of points significantly
outside the $99\%$-confidence lines. There were only three models in 16000
simulations with $\hat{C}_{181}$ lower than the observed value. Outlying
points also seem to come in clusters, with the binned power spectrum
having some wiggles\cite{Hinshaw:2003ex}.

\subsection{Goodness of fit}

The goodness of fit of the observed points to the theoretical model
can be assessed by using some kind of effective $\chieff$
value. Ref.~\refcite{Spergel:2003cb} reports an effective $\chieff=1431$
based on the likelihood of the theoretical best fit model given the
observed $\Chat$ estimators (including the T-E correlation power spectum). This may be surprisingly high for the 1342
degrees of freedom, indicating
that the standard best fit model has rather low probability given the
data. However we note a few technical points:
\begin{itemize}
\small
\item
The otherwise excellent third-order likelihood approximation used by
WMAP\cite{Verde:2003ey} cannot be expected to be accurate for points far from
the theoretical value. The values of the likelihoods for the outlier
points are probably underestimated and hence giving an
artificially high $\chieff$ (and could be skewing
parameter estimates, c.f. the section on the quadrupole below.)
\item 
Noise contributions to the variance of the $\Chat$ have been neglected
at $\ell < 100$ even though the noise at $\ell\alt 100$ is about five
times larger than for $\ell \agt 100$ because fewer frequencies are
being used. This overestimates the effective $\chieff$ by about 16 (though has
no significant effect on parameter estimates).
\end{itemize}

The best way to asses whether an effective $\chieff$ is acceptable is
by simulations\footnote{My simulations are rather crude, using single
  maps and modelling the noise as an isotropic contribution to the
  variance, which may give wider dispersion than a full simulation. I
  take the best-fit theoretical model as fixed.
 My $\chieff$ however accounts for the noise at $\ell<100$.}. 
Taking the temperature data only at $\ell < 250$ the
observed $\chieff$ is $275$, compared to simulations which give
$248 \pm 25 (1 \sigma)$. Thus on these scales where the noise is small there is not strong
evidence that anything is amiss. More detailed simulations would be
needed on smaller scales to assess whether the $\chieff=963$ over $\ell<900$
is acceptable or not. See also the section below on the
temperature-polarization cross-correlation.


\subsection{Low quadrupole}

The low value of the quadrupole has attracted considerable
attention. Although the likelihood of the observed $\hat{C}_2$ is not
that low for standard models (just outside the the 68\% likelihood contour in
Fig.~\ref{Cl}), it is atypical in being low. This means that any model
which predicts a low value for the quadrupole will have a higher
posterior probability than standard models, because in the new model
the low value can be much more typical (and so have much higher
likelihood). Some points about the quadrupole value:
\begin{itemize}
\small
\item
The very low value reported by WMAP of $\hat{C}_2=123\muK^2$ has unmodelled 
foreground uncertainties. For example treating the foregrounds differently by
using the map from Ref.~\refcite{Tegmark:2003ve} one obtains a
pseudo-$\Cl$ estimate $\hat{C}_2 = 184\muK^2$. (See also Ref.~\refcite{Efstathiou:2003dj}.)
\item
Ref.~\refcite{Tegmark:2003ve} claim that the value is low partly
because of a coincidence that the galaxy obscures regions of high
power. This appears to be a much smaller effect than the foreground uncertainty, as cutting the galactic region
from the map used in that paper only lowers the $\hat{C}_2$ by $\sim 10\%$.
\item
The exact value of the estimator and the likelihood value for
different models is
sensitive to how accurately you model the distribution. 
There are significant differences between a near-optimal analysis using
orthogonalized harmonics\cite{Gorski94,Mortlock00} and the WMAP
pseudo-$\Cl$ and likelihood parameterization.\footnote{
See the plot at \sf{http://cosmologist.info/Sorrento03}.
Note that problems of modelling the quadrupole likelihood do not
appear to significantly bias parameter constraints for different
initial power spectrum models. Even though
absolute likelihood values may differ over the range of models for
different methods, the
models always predict much higher values than the observed value and the slope of the
likelihood function is approximately correct over the relevant range.
} 
\end{itemize}

Without including Lyman-$\alpha$ data on very small scales, some weak
evidence for running of the spectral index arises from the low
value of the WMAP quadrupole\cite{Bridle:2003sa} favouring low
theoretical quadrupole values. Other models which favour low
quadrupoles are also favoured, though very low values cannot be
obtained by only changing the initial power spectrum. This is because
about half of the power in the quadrupole comes from the Integrated
Sachs Wolfe (ISW) effect, which is sourced by changing potentials along the
line of site. Though the large scale power can be decreased to lower
the contribution to the quadrupole from last scattering, the ISW
contribution cannot be lowered without decreasing the power on smaller
scales which is inconsistent with the values of the other $\Cl$.
The ISW effect can be decreased by changing the dark energy model from
a cosmological constant\cite{Weller:2003hw}, but when combined with a
strange initial power spectrum this is becoming
a seriously contrived model.


\section{Temperature-polarization correlation}

\begin{figure}[ht]
\psfrag{l}[][][0.7]{$\ell$}
\psfrag{Cl}[][][0.7]{$(\ell+1) C_\ell/\mu \text{K}^2$}
\centerline{\psfig{figure=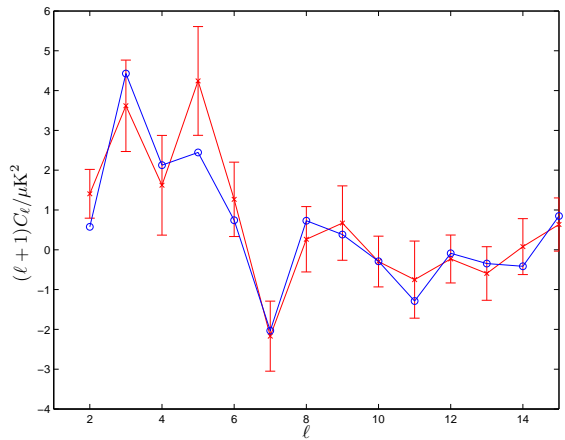, width=2.9in}}
\caption{
The temperature-polarization correlation power spectrum estimators at $\ell<16$ provided by WMAP. Error
bars are noise only, and both sets of points are available from the
WMAP web page.
\label{TE}}
\end{figure}
The power spectrum for the correlation between the temperature and
E-polarization appears to indicate a significant optical depth to last
scattering\cite{Kogut:2003et}. The relevant large scale power spectrum
is plotted in Fig.~\ref{TE}, showing the excess of power on large
scales that is a signature of reionization. The points with the error
bars are those plotted in Ref.~\refcite{Kogut:2003et}, whilst the
other set of points are the values actually used to compute the
likelihoods\footnote{\sf{http://lambda.gsfc.nasa.gov/}}\cite{Verde:2003ey}. Presumably these are different
analyses of the data, and the significant difference between the
points indicate a significant unaccounted for systematic error in the
   measurement of the TE power spectrum. It is interesting to note
   that $\ell=7$ is quite discrepant from theoretical expectations, but
   does not change significantly between the two analyses. The more
   than factor of two systematic difference in the TE quadrupole values likely
   explains the apparent inconsistency with the temperature quadrupole value 
   discussed in Ref.~\refcite{Dore:2003wp}. The WMAP likelihood
   analysis also neglects the correlation between the
temperature and TE power spectra, though this does not have a large
effect on joint-constraint cosmological parameter values.\footnote{Including the correlation is no harder than ignoring it at Gaussian order, though it is not
trivial to find a third-order likelihood parameterization accounting
for the correlations. One can compute the
likelihood distributions essentially exactly on large scales using
orthogonalized harmonics\cite{Mortlock00} (or Monte-Carlo
methods\cite{Wandelt:2003uk}) with Gaussianity assumptions, which
should probably at least be used as a check on any parameterization
that is used when better (and more foreground-free) data is
available.} 

\section{Asymmetries}

One of the main theoretical assumptions of most cosmological models is that the
universe is statistically isotropic: after removing the local dipole
the CMB anisotropy should not have any alignment other than random variations
expected from throwing down realizations of statistically isotropic
random fields on the sky. Ref.~\refcite{deOliveira-Costa:2003pu} find
in fact that the quadrupole and octopole have some rather unlikely
alignment properties, with the octopole being nearly planar, and the
axes of the two being closely aligned (however a posteriori statistics
for small samples of numbers should be interpreted with a little
care). 

Interestingly Ref.~\refcite{Eriksen:2003db} report that there
are also highly significant power anisotropies on larger scales. As an
independent minor variation on this theme I computed a single binned
$\Chat$ with $\ell\le 30$ over half the sky as a function of the axis of the
hemisphere, trying 48 different orientations. There is an apparent
asymmetry with a hemisphere centred close to the north ecliptic pole
giving about 35\%
less power than the opposite hemisphere. The significance of this
result is readily checked by simulating skies according to the
predictions of some Gaussian $\Lambda$CDM model, and computing the
maximum value of the power ratio in the different realizations. The observed value
lies 2---3 sigma away from the maximum value of the power ratio,
indicating that such a large power deficit over the northern ecliptic
hemisphere is rather unlikely. The axis of asymmetry can also be seen
(though with less statistical significance) for smaller bins in
$\ell$, though does not persist to small scales. It appears that there
is a region of startlingly low large scale power somewhere around the
north ecliptic pole. Colour figures are
available on the web\footnote{\sf{http://cosmologist.info/Sorrento03}}.

\section{Conclusions}

It would appear that the WMAP data might be inconsistent with simple
isotropic Gaussian cosmological models. Given the past experience
(e.g. with BOOMERANG) one's first suspicion naturally falls on the
data, and the alignment of power with the ecliptic (the axis of
symmetry of the observation) may be a hint in
this direction. However the features appear to be quite robust, and
deficits in power are quite hard to explain with foregrounds. One
possible explanation for the outlying points in the power spectrum might
be non-Gaussianity, with $a_{lm}$s being more likely close to zero or
large than if they had a Gaussian distribution. 

Whatever the resolution of
the puzzle, the current determination of cosmological parameters
assuming everything fits our assumptions is potentially misleading. A
small number of outlier points with large weight under the Gaussianity
assumption could be skewing
parameter estimates if they are included in the analysis but in fact have
their origin in some other systematic or non-Gaussian physics.
However the concordance of parameter estimates does suggest that this
effect may be
small. In any case it should be a matter of some priority to check the
WMAP results. 

Ref.~\refcite{Wandelt:2003uk} has presented a very nice Monte Carlo
method 
for computing cosmological parameter likelihoods essentially exactly if the  CMB really is
Gaussian and isotropic. This can include
foreground uncertainties consistently, and avoids all the problems
with computing accurate likelihood values from a set of $\Chat$ estimators.

\section*{Acknowledgments}
I thank the organisers, in particular Claudio Rubano, for inviting me
and hospitality. I thank Sarah Bridle and Matias
Zaldarriaga for useful discussions, and also George Efstathiou, Licia
Verde, Hiranya Peiris and Gary Hinshaw for communication about the
WMAP data.
I acknowledge the use of the Legacy Archive for 
Microwave Background Data Analysis (LAMBDA). Support for LAMBDA is
provided by the
 NASA Office of Space Science. The beowulf computer I used was funded by the Canada 
Foundation for Innovation and the Ontario Innovation Trust.

\appendix


\end{document}